\documentclass[prl,preprint,nofootinbib,preprintnumbers,amssymb,amsfonts,amsmath,superscriptaddress,showpacs,hyperreft]{revtex4-1}

\usepackage{amsmath,array,amssymb}
\usepackage{graphicx}

\usepackage{float}

 \usepackage{braket}
    \DeclareMathOperator{\sech}{sech}
\usepackage{color}

\begin{document}

\title{Spontaneous formation of a macroscopically extended coherent state }%

\author{C. Braggio}
\email{caterina.braggio@unipd.it}
\affiliation{Dip. di Fisica e Astronomia, Universit\`a di Padova and INFN, Via F. Marzolo 8, I-35131 Padova, Italy}

\author{F. Chiossi}
\affiliation{Dip. di Fisica e Astronomia, Universit\`a di Padova and INFN, Via F. Marzolo 8, I-35131 Padova, Italy}

\author{G. Carugno}
\affiliation{Dip. di Fisica e Astronomia, Universit\`a di Padova and INFN, Via F. Marzolo 8, I-35131 Padova, Italy}

\author{A. Ortolan}
\affiliation{INFN, Laboratori Nazionali di Legnaro, Viale dell'Universit\`a 2, I-35020 Legnaro, Italy}

\author{G. Ruoso}
\affiliation{INFN, Laboratori Nazionali di Legnaro, Viale dell'Universit\`a 2, I-35020 Legnaro, Italy}

\begin{abstract}
\bf It is a straightforward result of electromagnetism that dipole oscillators radiate more strongly when they are synchronized, and that if there are $N$ dipoles, the overall emitted intensity scales with $N^2$. In atomic physics, such an enhanced radiative property appears when coherence among two-level identical atoms is established, and is well-known as \lq\lq superradiance' \cite{Dicke:1954aa}. In superfluorescence (SF), atomic coherence develops via a self-organisation process stemming from the common radiated field, starting from a incoherently prepared population inversion \cite{Bonifacio:1975aa}.
First demonstrated in a gas \cite{Skribanowitz:1973} and later in condensed matter systems \cite{Florian:1984}, its potential is currently being investigated in the fields of ultranarrow linewidth laser development for fundamental tests in physics \cite{Meiser:2009,Meiser:2010,Bohnet:2012aa,Norcia:2016, Norcia:2018}, and for the development of devices enabling entangled multi-photon quantum light sources \cite{Raino:2018aa,Angerer:2018aa}. 

A barely developed aspect in superradiance is related to the properties of the dipole array that generates the pulsed radiation field. In this work we establish the experimental conditions for formation of a macroscopic dipole via superfluorescence, involving the remarkable number of $4\times10^{12}$ atoms. Even though rapidly evolving in time, it represents a flexible test-bed in quantum optics. Self-driven atom dynamics, without the mediation of cavity QED nor quantum dots or quantum well structures, is observed in a cryogenically-cooled rare-earth doped material.
We present clear evidence of a decay rate that is enhanced by more than 1-million times compared to that of independently emitting atoms. We thoroughly resolve the dynamics by directly measuring the intensity of the emitted radiation as a function of time.

\end{abstract}

\maketitle
{\em Introduction.}--- 
The paradigm effect for collective behaviour in quantum optics is superradiance (SR), extensively studied both theoretically and experimentally  starting from its prediction by Dicke in 1954  \cite{Dicke:1954aa}. 
A strong radiative coupling is established among $N$ excited atoms that are indistinguishable owing to their being confined within a small volume $V< \lambda^3$, where $\lambda$ is the radiated field wavelength. 
 The correlation among atoms gives rise to a macroscopic dipole, whereby each individual dipole accelerates the transition rate of the other emitters according to the  \lq \lq Dicke ladder'' representation.
Signatures of this effect are then to be sought in the natural $N$-squared scaling of the emission intensity by an array of phase-locked identical dipoles $N$ or in the temporal dynamics of their emission, described by sech-squared shape, intense coherent photon pulses. 

An extension of the SR concept to much larger volumes ($V\gg \lambda^3$) has been later theoretically analysed \cite{Rehler:1971aa}, thus implying the possibility of experimentally accomplishing a proportionally larger number of atoms $N$.
 
A cooperative spontaneous emission (i.e. radiation rate proportional to $N^2$) can also be obtained from an atomic system initially excited with zero macroscopic dipole moment, as is the case in superfluorescence (SF) \cite{Bonifacio:1975aa}.
Such an enhanced radiative property is hard to observe in condensed matter compared to atomic and molecular gases (see Ref.\,\cite{Cong:2016aa} and references therein) due to stringent requirements: that the atoms be indistinguishable, decoupled from their environment and that their density is high. In terms of spectroscopic properties of the material these demands translate to: small inhomogeneous line-broadening, small atomic dephasing rate and large enough inversion density. 
In the optical regime strong light-matter coupling via SF has been recently accomplished in nanostructured materials \cite{Raino:2018aa,Cong:2016aa,Laurent:2015,Timothy-Noe-II:2012aa,Jho:2006aa}, involving small atom number (up to 100) and short collective state lifetime ($\sim$\,ps).  

Our work pushes into a new regime, demonstrating spontaneous formation of a macroscopic dipole composed of the remarkable number of atoms $N\simeq4\times10^{12}$ in erbium-doped yttrium orthosilicate (Er:YSO, Er$^{3+}$:Y$_2$SiO$_5$). This optical material exhibits the narrowest homogeneous linewidths and the longest coherence lifetimes \cite{Thiel:2011aa}, is widely investigated for spectral hole burning applications \cite{Macfarlane:1997aa,Rancic:2017aa}, cavity QED \cite{Gonzalvo:2015,Probst:2014}, and the reversible, coherent conversion of microwave photons into the optical telecom C band around 1.54\,$\mu$m \cite{Probst:2015}. 
Prior to our work, superfluorescence from bulk crystals had been demonstrated for O$^-_{2}$ centres in an alkali halide crystal \cite{Florian:1984,Malcuit:1987ab}, even though the reported pulsed emission is clearly a superposition of pulses generated by small subsystems of correlated molecules \cite{Ishikawa:2016aa}. 
A further interesting aspect of our giant dipole is its lifetime, exceeding by several orders of magnitude previously reported macroscopic coherence times. Spontaneous coherence has in fact a characteristic induction time in superfluorescence \cite{Malcuit:1987ab}, typically comparable with $(10-100\,)$ times the characteristic SF emission rate, and we can infer a lifetime $\lesssim 1\,\mu$s from the recorded 50\,ns-duration coherent pulses. A physical system that exhibits superradiance is in principle capable to superabsorb photons \cite{Higgins:2014aa}, even though under normal conditions the coherent atomic ensemble is prone to radiate rather to absorb. To alter this tendency, a ring-like dipole arrangement, inspired to natural photosynthetic complexes, has been proposed in the framework of quantum nanotechnology, and through analytical and numerical calculations its potential for observation of the searched effect has been demonstrated \cite{Higgins:2014aa,Brown:2019}, even though yet experimentally elusive. Because of its long lifetime and high atom number, the macrocoherent transient state demonstrated in the present work could actively be probed with an intense laser field, with a non vanishing probability of superabsorption to an excited state.

Beyond its intrinsic interest and its potential applicability to energy harvesting, a superabsorber is sensitive to low microwave and light levels, a desirable feature in the context of quantum sensors development for future scientific instruments  \cite{Higgins:2014aa,Brown:2019}. Furthermore, the possibility to investigate superabsorption could revolutionise the detection of elusive particles  whereby a N-fold enhancement factor of event rates might be accomplished in upconversion schemes that have been recently proposed \cite{Braggio:2017aa,Sikivie:2014}.

{\em Experimental.}--- 
 \begin{figure}[h!]
\includegraphics[width=0.8\textwidth]{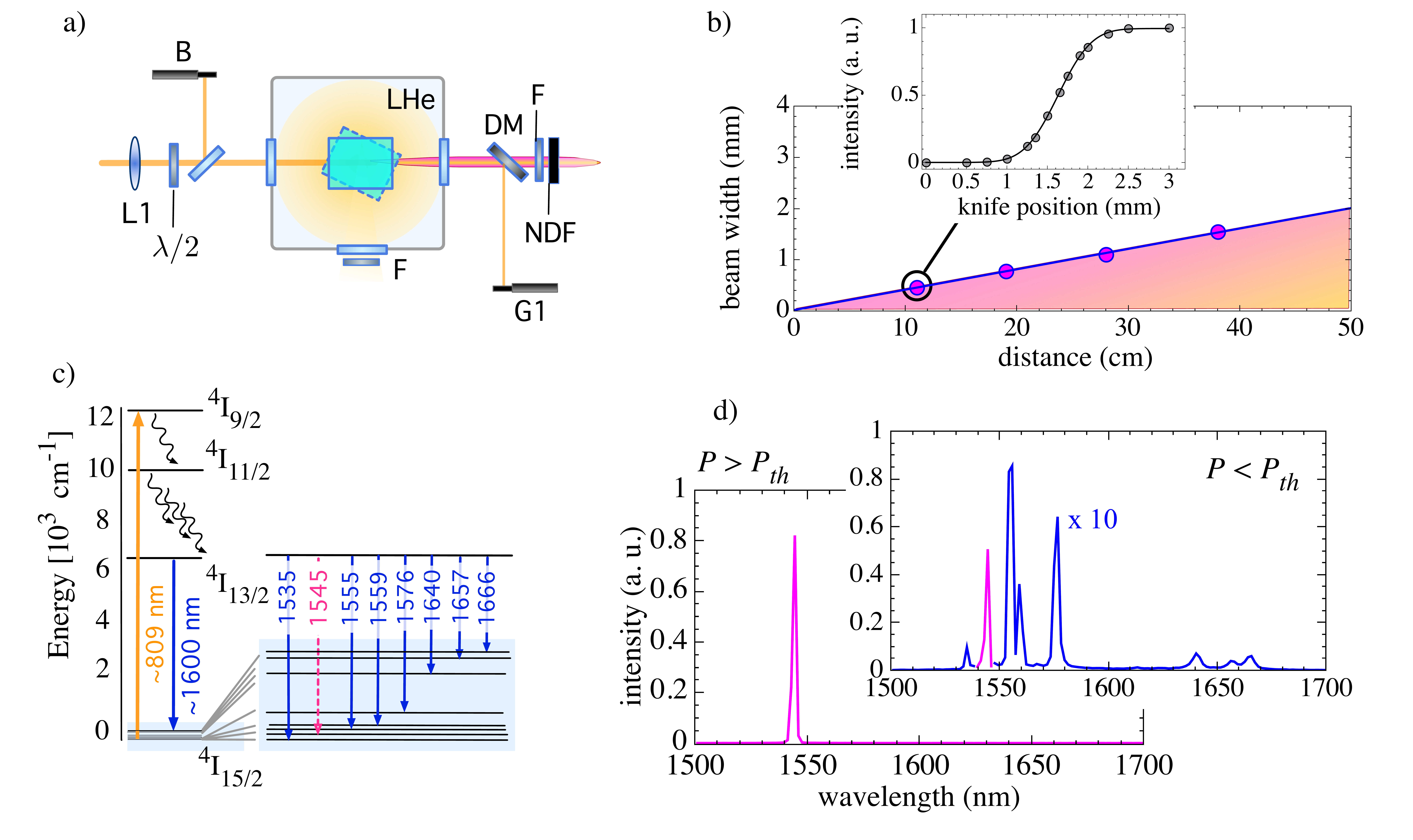}
\caption{(a) Experimental setup. A 0.1\% concentration, $4\times 5 \times 6.2$\,mm$^3$-volume Er:YSO crystal is immersed in LHe and maintained at 1.6\,K. Er ions are indirectly pumped with a Ti:sapphire laser (stabilized ring cavity, 10\,MHz linewidth) to the metastable level $^{4}$I$_{13/2}$ shown in (c). The formation of the macroscopic dipole is accompanied by coherent emission in the forward direction. The incident laser beam is focused by lens L1 and its polarisation can be rotated by means of half-wave plate $\lambda/2$. Dichroic Mirror DM reflects the pump laser light towards sensor G1 allowing for contextual laser absorption measurements. The  forward light transmitted through DM is attenuated by a set of calibrated neutral density filters (NDF). Longpass filters (F) are used to remove stray laser light. 
(b) Superfluorescent beam propagation and diffraction in the pencil-shaped sample geometry. The beam divergence angle $\theta_d$ is obtained by measuring with the knife-edge technique the forward beam radius at several distances from the crystal centre, and the obtained $\theta_d$ value is compatible with diffraction from an aperture whose radius coincides with the measured laser beam waist $\omega_0=163\,\mu$m at the crystal position ($\pm 1$\,cm). (c)-(d) Emission spectral filtering. Above the threshold $P_{th}$, superfluorescence emission takes place through the highest branching ratio transition, represented in the Er$^{3+}$ energy levels scheme by the dotted line arrow. 
}
\label{Fig:1}
\end{figure} 
Atoms that can participate to the spontaneous formation of the macroscopic dipole are Er$^{3+}$ ions, 
positioned at about 5\,nm relative distance (corresponding to 0.1\% atomic percent substitution for Y$^{3+}$) in a YSO crystalline host matrix cooled to 1.6\,K. Site 1 \cite{Bottger:2006} Er ions are incoherently excited by a cw pump laser to the long-lived ($\tau_0=15.0\pm0.1$\,ms) $^4$I$_{13/2}$ level via phonon-emitting steps as shown in Fig.\,\ref{Fig:1} (a). 
The atomic sample is then automatically endowed with the shape of a long cylinder of length $L$, where $L=6.2$\,mm is the crystal length along the laser propagation direction, and transverse dimension $2\omega_0$ ($\omega_0$ laser beam waist), obeying the relation $L> \omega_0 \gg \lambda$ (pencil-shaped sample), with $\lambda$ wavelength of the SF beam. 

Clear evidence of the formation of a macroscopic dipole in our physical system is the observation of pulsed superradiant emission above a well defined threshold value of inversion population density. When this condition is satisfied, we record a bright forward field whose emission spectrum displays spectral filtering, e.g. the natural multiline emission spectrum collapses to the highest branching ratio transition at $\lambda = 1.545\,\mu$m (Fig.\,\ref{Fig:1} (d)). 

Knife-edge measurements (Fig.\,\ref{Fig:1} (c)) show that the beam at $1.545\,\mu$m-wavelength emerging from the crystal is a gaussian beam with far-field beam divergence $\theta_d=3.6\pm0.5$\,mrad. This value is comparable with the calculated diffraction angle $\theta_D=\lambda/2\omega_0=4.7$\,mrad of the pencil-shaped atomic sample, with $\omega_0=163\,\mu$m measured pump beam waist.
 Such a remarkable control of the SF beam parameters is crucial to establish the role of diffraction in superradiance processes, clearly distinguishing emission supported by off-axial modes from diffraction of each mode propagating close to the cylinder axis \cite{Heinzen:1985,Mattar:1981,Benedict:1996}. The Fresnel factor $F= \pi \omega_0^2/L\lambda$ is the key parameter to set the conditions to enter such regimes giving a F-lobes pattern for $F>1$ and a single, wide-area lobe when $F<1$  \cite{haroche:1982}. In the present configuration $F\sim 9$, with emission intensity mainly concentrated within a single lobe as wide as $\theta_D$. 
 
  {\em Observation of superfluorescent pulses.}--- \label{sfd}
 
The described beam propagation study allows proper optical coupling of the SF beam profile to small-area, ultrafast photodiodes to investigate the dynamics. In the following we demonstrate that the pencil-shaped sample geometry determines not only the strong directionality of the SF light beam, but influences the temporal dynamics of the corresponding photon bunches. In fact, it mitigates the characteristic SF emission time $\tau_{\mbox{\tiny R}}$ through the geometrical factor $\mu=3\Omega_0/(8\pi$), proportional to the diffraction solid angle $\Omega_0=\lambda^2/(\pi\omega_0^2)$ of the sample
\begin{equation}
\label{tauR}
\tau_{\mbox{\tiny R}}=\bar{\tau}/\mu N,
\end{equation}
where $\bar{\tau}=\tau_{0}\beta$ is the inverse of the transition rate at 1545\,nm wavelength (Fig.\,\ref{Fig:1} (c)), and $\beta=2.36$ is estimated by the spectra recorded below SF threshold (Fig.\,\ref{Fig:1} (d)).  

The SF full dynamics is enclosed in the hyperbolic secant shape of the emitted intensity \cite{haroche:1982}
\begin{equation}
I(t)=\frac{hc}{\lambda}R_p\sech^{2}[(t-t_0)/2 \tau_{\mbox{\tiny R}}],
\label{eq:I}
\end{equation}
where $R_p$ is the peak photon output rate (i.e. pulse amplitude at $t=t_0$), proportional to $N^2$ as expected for SR. 

\begin{figure}[h!]
\includegraphics[width=0.5\textwidth]{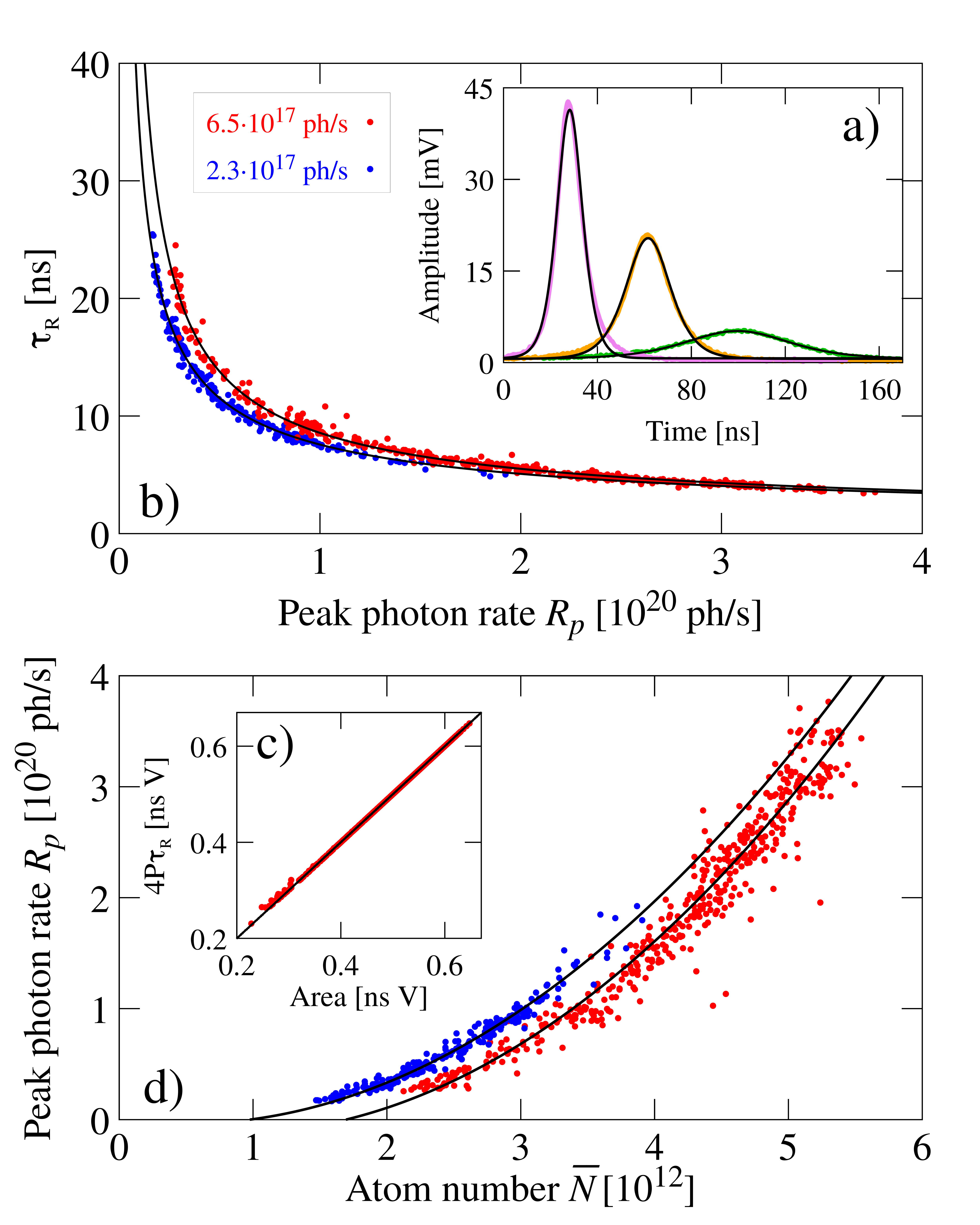}
\caption{Spontaneously generated superfluorescent pulses at different pump fluences. (a) Representative individual time traces of photon bunches recorded under identical excitation conditions. The time dependence of the signal amplitude, proportional to the average photon emission rate, is very well fitted by a squared hyperbolic secant function. 
(b) Pulse duration $\tau_{\mbox{\tiny R}}$ versus peak photon rate $R_{p}$ for pumping rates. The black lines are fits in the form  given by eq.\,\ref{eq:II}, giving an average $\mu=(2.4\pm0.1)\times10^{-6}$, compatible with diffraction from an aperture of diameter $2\omega_0=326\,\mu$m.
(c) Nominal area of the fit function given by eq.\,\ref{eq:I} versus integrated pulse area. Linear regression with unitary slope demonstrates pure SF dynamics for each recorded photon bunch.
  (d) Peak photon rate versus observed atom number. The recorded data scale superlinearly, as shown by the $N^2$-trend represented by the black line.}
\label{Fig:2}
\end{figure}

We predict that, when the single pass gain $\alpha L$ of the inverted medium is high, eq.\,\ref{eq:I} is modified to 
\begin{equation}
I(t)=\frac{hc}{\lambda}\frac{\mu\bar{N}(\bar{N}-N_0)}{4\bar{\tau}}\sech^{2}[(t-t_0)/2 \tau_{\mbox{\tiny R}}],
\label{eq:II}
\end{equation}
having introduced the parameter $\bar{N}=N(1+\alpha L)=N+N_0=4\tau_R R_p$. As one photon in the bunch corresponds to one atom participating to the coherent process, $\bar{N}$ represents the number of experimentally observed photons, in excess of those truly emitted by the macroscopic dipole owing to the medium gain. 
Figure\,\ref{Fig:2}\,(a) shows three representative photon bunches, selected from several hundreds that have been analysed to prepare the main plots for given pump laser fluences, revealing the stochastic nature of the effect \cite{Vrehen:1980aa,Vrehen:1981aa,Florian:1984}. All the recorded pulses are well-fitted by a pure sech-squared temporal profile, as confirmed by the linear regression in Fig.\,\ref{Fig:2}\,(c).

From Eq.\,\ref{eq:I} and \ref{eq:II} we obtain: 
\begin{equation}
R_p = \frac{hc}{\lambda}\frac{\mu \bar{N} (\bar{N}-N_0)}{4\bar{\tau}}, 
\label{eq:IIa}
\end{equation}
in which the superlinear scaling of the peak photon rate with $\bar{N}$ is explicit. 
By recalling that $\tau_R = \bar{\tau}/[\mu (\bar{N}-N_0)]$ and using Eq.\ref{eq:IIa}, the superfluorescent time can be written as:
\begin{equation}
 \tau_{\mbox{\tiny R}}=\frac{N_0}{8\,R_p}\left[1+\sqrt{1+16R_p\frac{\bar{\tau}}{\mu N_0^2}}\right],
\label{eq:III}
\end{equation}
The equations \ref{eq:IIa} and \ref{eq:III}  connect the three observables $\bar{N}, \tau_R$ and $R_p$, and accurately fit the recorded data as shown in Fig.\,\ref{Fig:2}, confirming the occurrence of superfluorescence amplified by stimulated emission. 

It is important to note that, owing to the parametrisation that we introduced to account for the medium gain, the recorded data are very well fitted by functional forms including two physical parameters, namely the emission gain $N_0$ and the geometrical factor $\mu$. Using an absolute calibration, we can estimate  $\mu$ directly by the fitting procedure as $(2.4\pm0.1)\times10^{-6}$, a value that favourably compares with that calculated by its definition $\mu=\frac{3}{8\pi^2}\left(\frac{\lambda}{\omega_0}\right)^2=3.4 \times 10^{-6}$, given the one-dimensional approximation \cite{haroche:1982}.

Excess photon values $N_0$ of $(1.0\pm0.1)\times 10^{12}$ and $(1.7\pm0.1)\times 10^{12}$ are obtained for the two data series at different fluence, accordingly to the greater gain in the higher inversion medium. Note that data at higher fluence  shown in Fig.\,\ref{Fig:2}\,(d) are characterised by more pronounced deviations as one expects when intrinsic fluctuations in SF get amplified by a greater gain medium. 

The most intense pulses in Fig.\,\ref{Fig:2} (d) definitely demonstrate macro-coherence involving more than 4$\times 10^{12}$ atoms.

{\em SF Emission average intensity.}
 
 According to $\tau_{\mbox{\tiny R}} \ll T_2^*$, the observed (4-52)\,ns-duration $\tau_{\mbox{\tiny R}}$ values reflect a SF transition linewidth much narrower than 10\,MHz. Incidentally, this latter value is comparable with the linewidth of our pump laser, reflecting to some extent the excited ions distribution width. By laser fluence absorption measurements we estimate a steady-state inverted atom number $N_1\gtrsim 10^{15}$ in the $^{4}$I$_{13/2}$ level, and it can reasonably be assumed that each SF photon bunch, arising from a subsystem of $\sim10^{12}$ ions as demonstrated by the plots in Fig.\,\ref{Fig:2} (d), does not significantly alter the excited level population density. In addition, the pulse repetition rate is $\sim 100$\,kHz, it is then possible in our physical system to report the SF intensity averaged over several thousands of independent pulses as a function of $N_1$. 

  \begin{figure}[h!]
  \includegraphics[width=\textwidth]{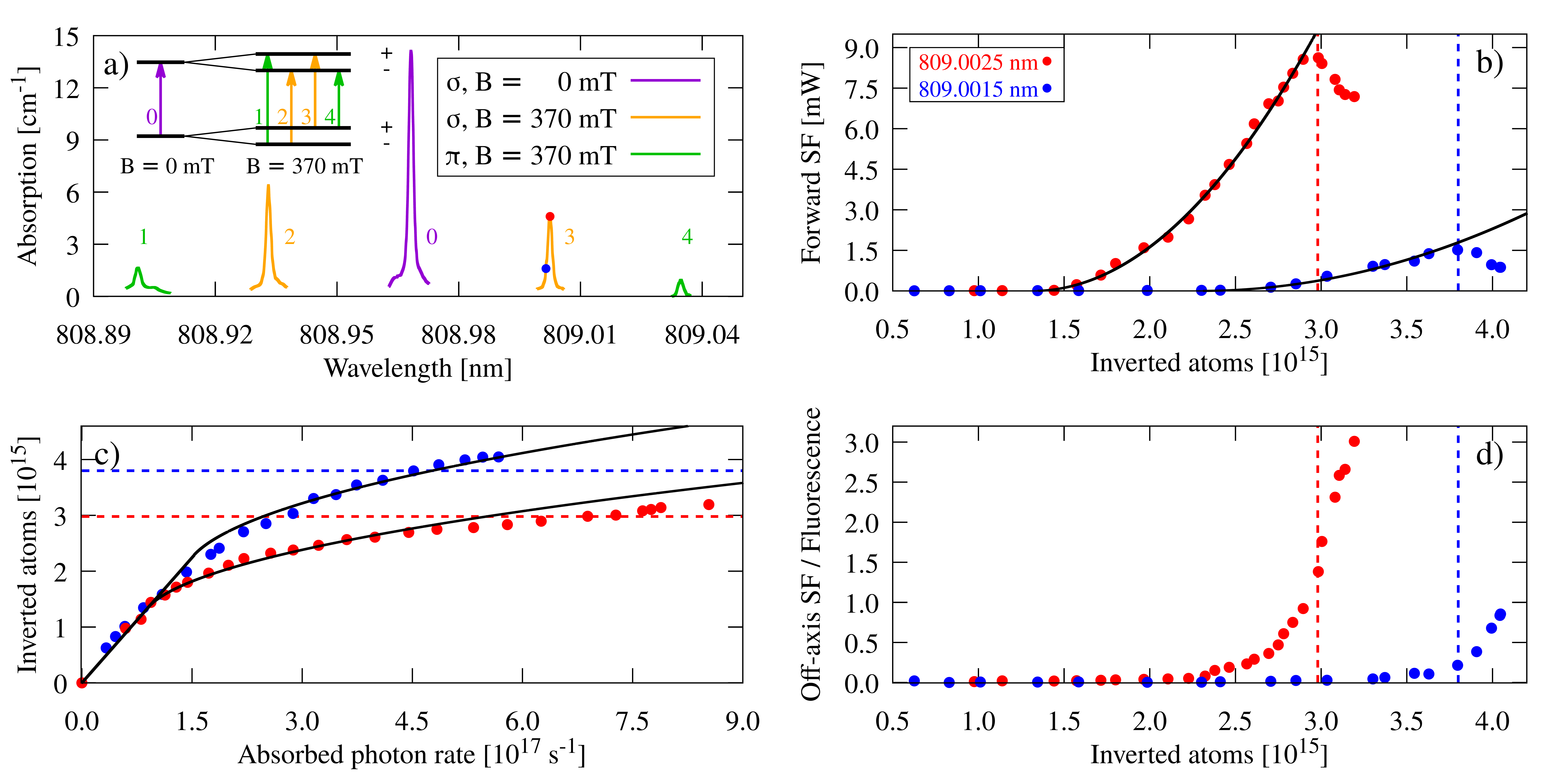}
  \label{Fig:3}
  \caption{(a) Absorption laser spectroscopy for the transition $^4$I$_{15/2} \rightarrow ^4$I$_{9/2}$ for $\pi$ and $\sigma$ polarization with and without 370\,mT magnetic field. (b) Forward (d) and off-axis average intensity of the superfluorescent emission for different longitudinal profiles of the inverted population density at absorption coefficients $1.6$\,cm$^{-1}$ and $4.6$\,cm$^{-1}$, corresponding to centred ($809.0025$\,nm) and slightly off-resonance ($809.0015$\,nm) wavelengths respectively (blue and red dots in Fig.\,3\,(a)). 
 The dotted lines indicate when the off-axis emission is so important as to reverse the growth trend in the forward emission. The black lines are a parabolic fit to the data.
   (c) Data agreement with the steady-state rate equations analysis. Deviation from the initial linearity of the population inversion with pump absorption is due to the superfluorescence emission, which is more pronounced in the data recorded at resonance (red dot data).}
\end{figure}

As shown in Fig.\,3\,(a), a sharp inhomogeneous absorption line of 600\,MHz is observed on the transition between the lowest Stark levels of the 
$^4$I$_{15/2}$and $^4$I$_{9/2}$ manifolds. 
 Both the narrow linewidth and the large oscillator strength endow the Er:YSO crystal with a absorption coefficient of 14.1\,cm$^{-1}$ at resonance, and we use a 370\,mT magnetic field to mitigate the pump laser absorption and in turn establish a quasi-uniform inverted population density. Emission is thus investigated for two values of pump laser wavelength, centred and slightly off-resonance as indicated by the blue and red dots in Fig.\,3\,(a). 
 
 In Fig.\,3\,(b) the forward SF beam average power is directly measured with a Germanium power sensor, while in the horizontal axis we report the corresponding inverted atom number.
 The latter is estimated from the fluorescence spectra acquired at a InGaAs CCD-based spectrometer, collecting photons emitted orthogonally to the laser propagation direction.  
 At low laser pump power the inverted population density linearly depends on the measured absorbed laser power, enabling a calibration procedure (see the Methods section) that provides the absolute value of inverted atoms throughout the whole excitation range as shown in Fig.\,3\,(c).  It is worth noticing that, for increasing values of inverted population, the same phenomenon of spectral filtering, reported previously for the emitted radiation in the forward direction, is observed in the spectra of the orthogonal emission, indicating transverse SF photon bunching (see Methods). The temporal dynamics of these off-axis photon bunches satisfies eq.\,\ref{eq:II} and \ref{eq:III}, even though with a larger characteristic $\tau_{\mbox{\tiny R}}$ compared to the forward pulses due to the smaller initial atom number $N$. In Fig.\,3\,(d) we report the ratio of these coherent pulses average intensity (a single line in the spectrum) with the incoherent fluorescence emission (7+1 lines in the spectrum), having subtracted from the first the component due to the scattered SF forward intensity. 
We observe that the intensity of the forward emission is proportional to ($N_1-N_{10})^2$ (Fig.\,3\,(b)), where $N_{10}$ is the SF threshold atom number. Deviation from the predicted $N^2$-dependence at higher pumping levels is fully ascribed to off-axis, omnidirectional superfluorescence, as confirmed by the plot in Fig.\,3\,(d). An increasing fraction of the atomic sample is involved in this transverse cooperative emission to such an extent that the forward emission trend gets inverted in both the data series obtained for two different population inversion profiles (see Fig.\,3\,(b)).  
 This is a new aspect that is of importance for the present work aims, previously not reported in experiments performed in the pencil-shaped sample geometry \cite{Gross:1976,Gibbs:1977,Florian:1984}.

We indirectly analyse the efficiency of the cooperative emission process by reporting the inverted population vs absorbed photons as shown in Fig.\,3\,(c). As the $^4$I$_{9/2}$ and $^4$I$_{11/2}$ levels quickly relax via efficient multiphonon relaxation to the $^{4}$I$_{13/2}$, its level population number $N_1$  can be thoroughly described by a single, steady-state rate equation:

\begin{eqnarray}
\label{ssrq}
{\rm{\Phi_{abs}}}-\mathrm{N_1}/\tau_0 = 0 \qquad  &(\mathrm{N_1}<{\mathrm{N}}_{10})\\
\label{ssrq1}
{\rm{\Phi_{abs}}}-\mathrm{N_1}/\tau_0-b(\mathrm{N_1}-{\mathrm{N}}_{10})^2=0 \qquad &(\mathrm{N_1}>{\mathrm{N}}_{10})
\end{eqnarray}

in which the $b$ coefficient quantifies the level depletion by the cooperative process, $N_{10}$ is the SF threshold inferred from the plot in Fig.\,3\,(b) and ${\rm{\Phi_{abs}}}$ is the absorbed photon rate. The solution of eq.\,\ref{ssrq} predicts a linear dependence of the inverted population vs absorbed photons when $N_1$ is below the SF threshold $N_{10}$, and a marked $\sqrt{\rm{\Phi_{abs}}}$-trend otherwise. As long as the off-axis emission is not relevant, data reported in Fig.\,3\,(c) are well fitted by this solution, and the gap to the projected linear fluorescence value gives the fraction of ions that is diverted to SF, as high as 74\% over the photon bunch duration time scale.  An evolution of this rate model has successfully been applied to data obtained in a different material (Er:YLF), in which we observed cascaded SF emissions at 2.7\,$\mu$m and at $1.55\mu$m [Ref], seeded by a cw inversion in the $^4$I$_{11/2}$ level. The SF cascade occurs because in the YLF matrix the levels $^4$I$_{11/2}$ and $^4$I$_{13/2}$ have comparable lifetimes (7.8\,ms and 16.5\,ms measured at 1.6\,K, respectively) owing to its smaller phonon energy as compared to the YSO matrix (the maximum phonon energy is $\sim1000$\,cm$^{-1}$ in Er:YSO \cite{Zheng:2007aa} and $\sim440$\,cm$^{-1}$ in Er:YLF \cite{Zhang:1994aa}). 
 
  {\em Conclusions and outlook.} 
 We have reported unambiguous evidence of superfluorescence achieved by incoherently seeding a cw population inversion on the 1.54\,$\mu$m transition in Er:YSO. 
  The observed cleanest sech-squared light pulses positively differ from the complex superradiance dynamics observed by several groups, which are affected by ringing effects, asymmetry with long tails and multiple structures, thus showing that coherent ringing is not an intrinsic feature of SF in solid-state extended media, as has generally come to be believed \cite{Heinzen:1985,Greiner:2000}. 
  
  The peak photon rate $R_p$ is demonstrated to scale quadratically with atom number and the light pulses width follows the predicted $1/\sqrt{R_p}$ trend, as long as the high gain of the inverted medium is taken into account. 
 
  Superradiant pulses being emitted at a rate a million times faster than the incoherent spontaneous emission time have been reported, demonstrating self-driven atom dynamics with no cavity mediation nor interaction engineering to further complicate the description and limit the number of involved dipoles.   
  The key $N$-squared scaling of the emission, which distinguishes SF from simple laser emission (ASE), is also found when we integrate the energy of thousands forward photon bunches. Interestingly, we have registered SF pulses that propagate orthogonally to the cylinder-axis atomic sample, indicating adequate optical thickness for SF also in the transverse direction for our highest pumping levels ($\sim 0.3$\,W/cm$^2$). Thus far, omnidirectional emission had only been reported in a Rb vapor sample, even though as a result of four-wave mixing \cite{Lvovsky:1999}. 

 A quantitative approach to SF has been developed, based on a single, simple rate equation at the steady-state for the fluorescent level $^{4}$I$_{13/2}$, which includes a superfluorescence-related term weighting both the pulse rate and amplitude. 
Still, in the framework of ultra-narrow laser development, it has been emphasised how pulsed emission is not an intrinsic property of superradiance, and that operation in a continuous manner, with pump lasers applied to return the atoms to the excited state, is a viable solution to reach the mHz linewidth laser \cite{Meiser:2009,Meiser:2010,Bohnet:2012aa}. The results of our SF beam propagation study finally comply with the diffraction-based theoretical description given long ago by Gross and Haroche \cite{haroche:1982} about superfluorescence by an extended sample of atoms. 
This study is of importance for the design of a resonant repumping cavity. 
 
  From the properties of the emitted superradiant field, we infer the spontaneous formation of a macroscopically extended coherent state, whose key parameters, namely the number of involved atoms ($\sim 4\times 10^{12}$) and induction time ($\leqslant 1\,\mu$s), promise applicability as a test-bed for quantum optics effects.   With the numbers already at hand in this work, a $\sim1$\,\%-duty cycle coherent sensor is envisaged, capable at coherently amplifying the smallest of signals, with an overall intrinsic gain of $4\times 10^{12}$ \cite{Carlson:1980aa}.
 
\section{Methods}
\begin{figure}
\includegraphics[width=8.5 cm]{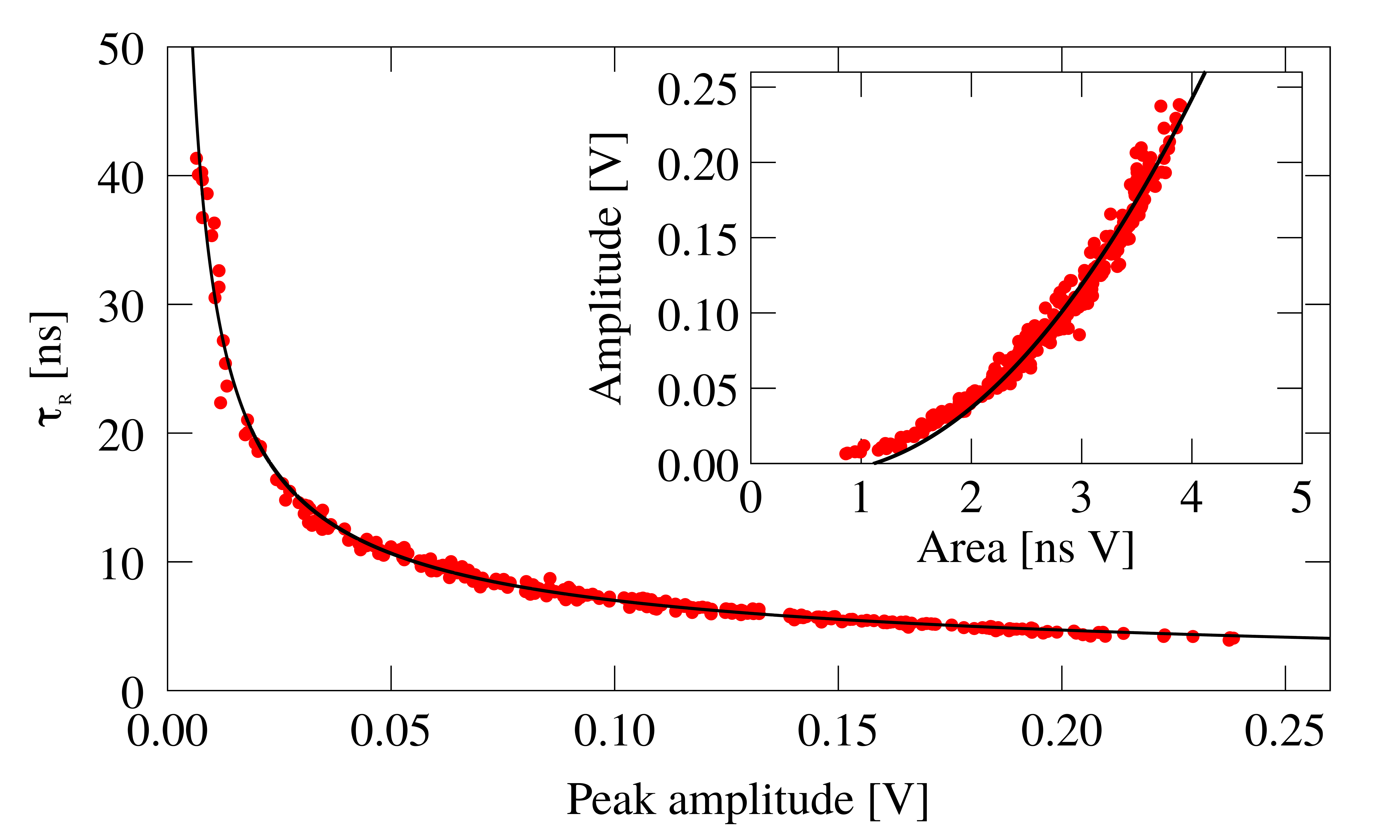}
\caption{Temporal dynamics recorded at an ultrafast InGaAs photodiode. }
\label{Fig:4}
\end{figure}

To operate the photodiodes in their linear response range the input forward SF light beam is attenuated by  
three neutral density filters, for a total transmittance of $2.3\times10^{-5}$. The SF beam is first collimated and then focused to a beam diameter compatible with full photon bunch detection at the fast photodiode area (10\,ns rise time, 0.8\,mm$^2$ area). Acquisition at a 6\,GHz digital sampling oscilloscope and analysis of the same pulses at ultrafast photodiode (25\,ps rise time, $10^{-3}$\,mm$^2$ area) ensured no distortion of the temporal dynamics recorded at the fast photodiodes as proven in Fig.\,\ref{Fig:4}.

The number of photons reported in the horizontal axis of Fig.\,\ref{Fig:2} (d) is thus obtained by computing the pulse area and taking into account the PD responsivity. 

{\em Calibration procedure}
An estimate of the inverted atom is obtained as follows. The integrated spectral intensity of the isotropic spontaneous emission (seven blue lines selected from the full spectrum reported in Fig.\,3\,(d)) is converted to inverted atoms number by imposing a linear dependence with slope $\tau_0$ to the data recorded at low pumping levels (up to absorbed photon rate $10^{17}$\,s$^{-1}$ in Fig.\,3\,(c)).
 
\begin{figure}
\includegraphics[width=8 cm]{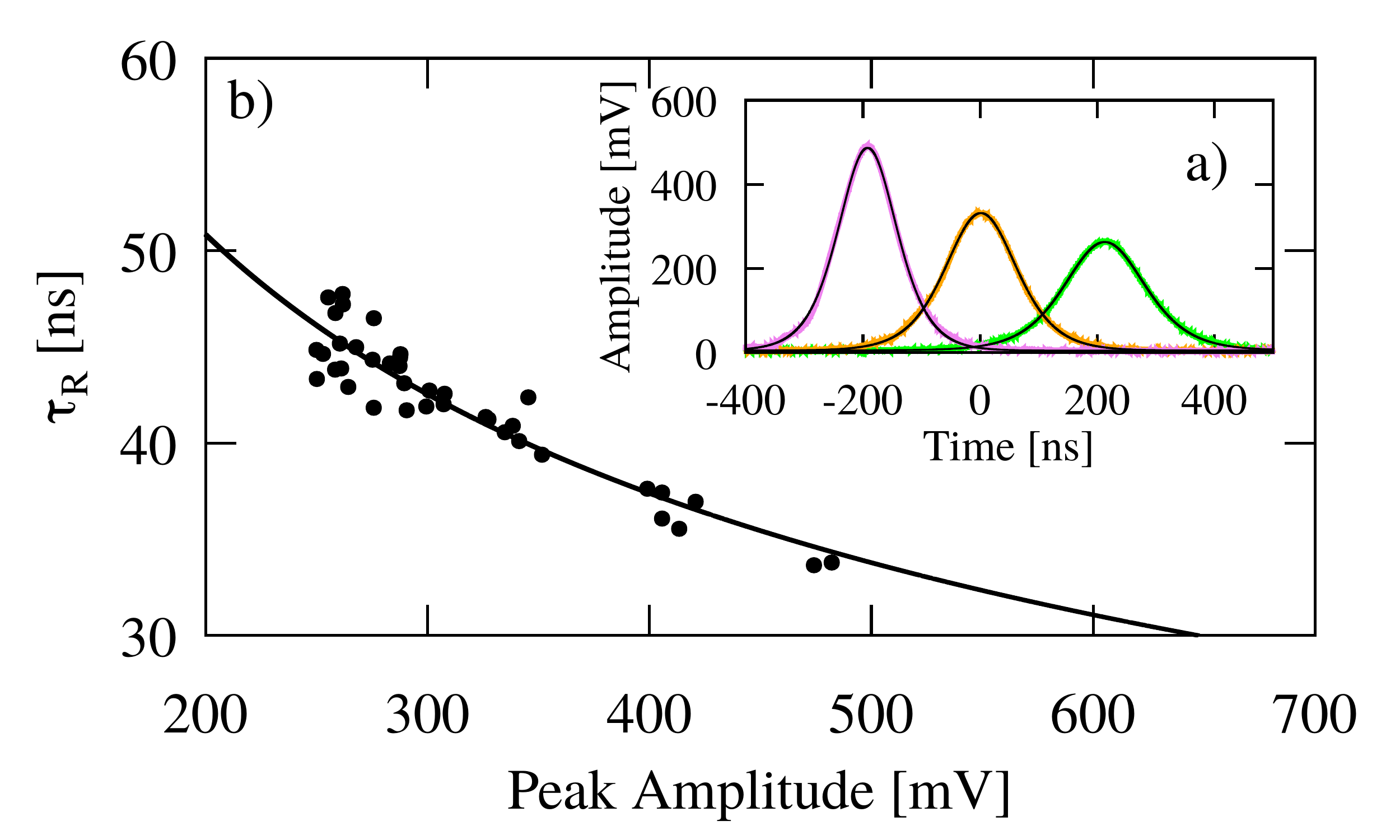}
\caption{Detection of transverse photon bunches.}
\label{Fig:5}
\end{figure}

{\em Transverse photon bunches}
The temporal dynamics of the transverse photon bunches has been investigated with a fast photodiode with no input coupling optics.  
Inset of Fig.\,\ref{Fig:5} shows three time-shifted representative photon bunches. The recorded transverse pulses are well-fitted by the same pure sech-squared temporal profile used for the forward bunches.

 \section{Author contributions}
 
 C. B and F. C. designed and performed the experiments. C. B. wrote the manuscript. F. C. analysed the data. G. C., A. O. and G. R. helped in revising the manuscript and contributed during the experiments. All authors discussed the results and commented on the manuscript.
 
 The authors declare that they have no competing interests.
 
  \section{Acknowledgements}
 We thank L.\,A. Lugiato for helpful discussions and critical reading of the manuscript.


 \bibliography{biblio3}

\end{document}